\documentclass[aip,reprint,a4paper]{revtex4-1}
\pdfoutput=1
\usepackage{graphicx}
\usepackage[utf8]{inputenc}
\begin{document}
\title{First measurements with the Munich 2D-ACAR spectrometer  on Cr}
\author{Hubert Ceeh}
\email{hubert.ceeh@frm2.tum.de}
\affiliation{Technische Universität München, Lehrstuhl E21, James-Franck Straße, 85747 Garching, Germany }
\author{Josef-Andreas Weber}
\affiliation{Technische Universität München, Lehrstuhl E21, James-Franck Straße, 85747 Garching, Germany }
\author{Michael Leitner}
\affiliation{Technische Universität München, Lehrstuhl E13, James-Franck Straße, 85747 Garching, Germany }
\author{Peter B\"oni}
\affiliation{Technische Universität München, Lehrstuhl E21, James-Franck Straße, 85747 Garching, Germany }
\author{Christoph Hugenschmidt}
\affiliation{Technische Universität München, Lehrstuhl E21, James-Franck Straße, 85747 Garching, Germany }
\affiliation{FRM\,II, Technische Universit\"at M\"unchen, Lichtenbergstra\ss e 1, 85747 Garching, Germany}
\date{\today}

\begin{abstract}
The Munich 2D-ACAR spectrometer at the  Maier-Leibnitz accelerator laboratory in Garching has recently become operational. In the present implementation a 2D-ACAR spectrometer is set up, with a baseline of $16.5\,$m, a conventional $^{22}$Na positron source and two Anger-type gamma-cameras. The positrons are guided onto the sample by a magnetic field  generated by a normal conducting electromagnet. The sample can be either cooled by a standard closed-cycle-cryostat to low temperatures  or heated by a resistive filament to temperatures up to $500\,$K.
We present the key features of this new 2D-ACAR spectrometer and, in addition, discuss first measurements on the pure metal system Cr. The 2D-ACAR measurements have been performed on Cr at different temperatures: at $5\,$K and at room temperature in the anti-ferromagnetic phase and at $318\,$K slightly above the paramagnetic phase transition.
\end{abstract}
\maketitle
\section{Introduction}
Almost all properties of a material are given by its electronic structure. For metallic materials, the most important characteristic of the electronic structure is the Fermi surface, namely the boundary between occupied and unoccupied states in momentum space.
The traditional experimental method for studying the Fermi surface is the de Haas-van Alphen effect, which is limited to low temperatures, and therefore cannot be utilized for the study of certain paramagnetic materials, for example Cr, since the Néel temperature of Cr is $311\,$K \cite{CrRev}. For this reason, numerous ACAR investigations, both experimental \cite{Shi,Sin,Fret} and theoretical \cite{Sint,Sun,Rub} have been performed on paramagnetic Cr. Studying the fermiology of the paramagnetic phase of Cr would allow to gain insight to the driving mechanism of the anti-ferromagnetic ordering of Cr below $311\,$K, since a nesting feature of the Fermi surface gives a possible explanation. However, several 2D-ACAR investigations performed in the paramagnetic \cite{Dug,Bia} phase of Cr  revealed significant discrepancies between the measured and the calculated two-photon-momentum distribution, which cannot be explained in terms of an unexpected topology of the Fermi surface. These discrepancies are most probably due to $e^+$-$e^-$-correlations and/or many-body effects not accounted for by theory. Recently, an experimental approach to this problem, the utilisation of a state-dependent enhancement has been presented in\cite{Lav}. With this method the enhancement is directly obtained by fitting the contribution of the individual bands to the data. This way, the Fermi surface nesting vector was deduced via the computation of the static susceptibility from the fitted LMTO calculations. In our study we focussed on the differences in the Fermi surface topology between paramagnetic and anti-ferromagnetic Cr. Therefore, 2D-ACAR measurements at different temperatures have been performed.

\section{Experimental setup}
The 2D-ACAR spectrometer consists of two main components, the source-sample chamber including the sample environment, and the detector assembly consisting of two Anger-type gamma cameras, which were obtained from the positron group of Bristol university. The Anger cameras are positioned symmetrically at a distance of $8.25\,$m to the source-sample chamber. The Munich 2D-ACAR spectrometer features two interchangeable sample holders, with an available temperature range between $5\,$K and $500\,$K, both of which were used for this study. The details of the Munich 2D-ACAR spectrometer and the performance characteristics can be found in \cite{Hub}.\\
For this study, four 2D-ACAR measurements have been conducted, each with the integration direction chosen along [100]. Two measurements were performed with the coolable sample holder in the anti-ferromagnetic phase of Cr, one at $5\,$K with a total of $1.10\cdot 10^8$ counts and the other one at room temperature with a total of $1.84\cdot 10^8$ counts. Two additional measurements were performed with the heatable sample holder, one at room temperature, in order to ensure reproducibility, with a total of $2.09\cdot 10^8$ counts and the other one at $318\,$K with a total of $2.63\cdot 10^8$ counts. Before each of the two measurement campaigns the orientation of the Cr crystal was checked by means of Laue diffractometry and aligned to  $\pm 1^{\circ}$. Before the sample was mounted the surface was etched with concentrated hydrochloric acid in order to remove the oxide layer from the surface. The crystal was cut by means of spark erosion, therefore it is assumed that the crystal is defect free. Positron lifetime measurements are underway to investigate the defect concentration. \\
The recorded ACAR spectra were corrected for the momentum sampling function and then folded back to \textbf{k}-space following the LCW-procedure\cite{LCW}. This way features in the projection of the momentum distribution attributed to the Fermi surface are enhanced and easier to identify. The LCW-folded data were corrected for  background contributions and scaled from $0$--$1$ so that the dynamic range is identical for all three measurements as well as for theoretical calculations: $0$ represents unoccupied states while $1.0$ represents the occupied states.\\
In addition, first principle band structure calculations have been performed for paramagnetic Cr using the Munich SPR-KKR-package \cite{Theo} and the expected two-photon momentum distribution was deduced. For the calculations the independent particle model was applied, therefore $e^+$--$e^-$--correlation effects are not included in the theoretical data.   
\section{Results}
\begin{figure}[h]
\begin{minipage}{0.43\textwidth}
\includegraphics[width=\textwidth]{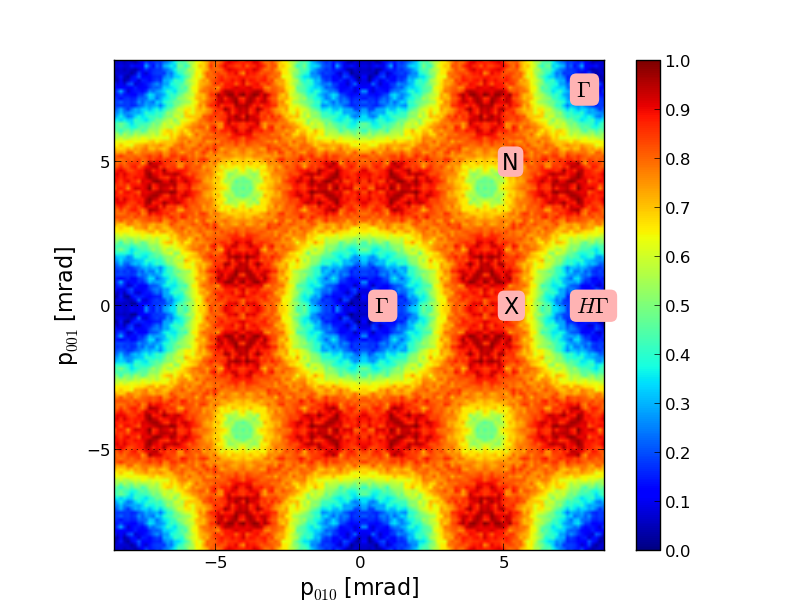}
\caption{\label{fig:room}LCW distribution for anti-ferromagnetic Cr [100] at room temperature.}
\end{minipage}\hspace{0.05\textwidth}%
\begin{minipage}{0.43\textwidth}
\includegraphics[width=\textwidth]{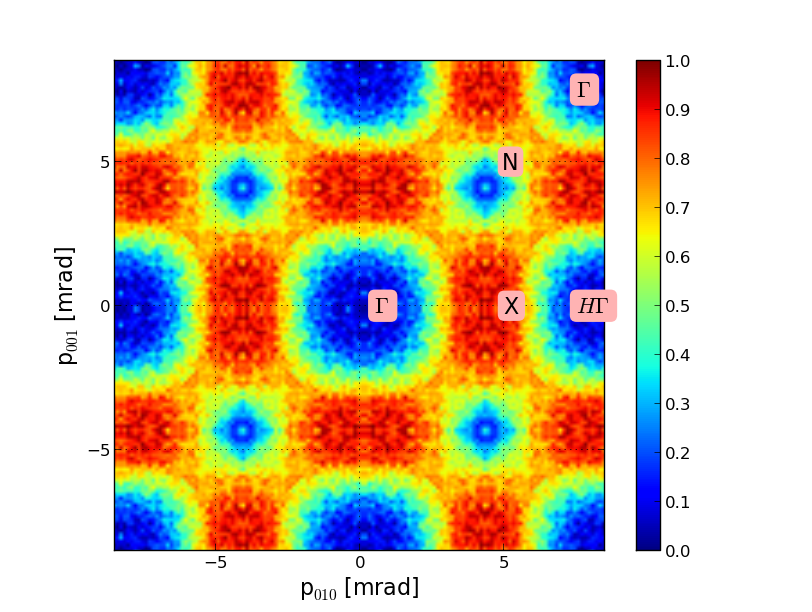}
\caption{\label{fig:318}LCW distribution for Cr [100] at $318\,$K. Slightly above the Néel temperature for Cr of $311\,$K.}
\end{minipage} 
\end{figure}
\begin{figure}[h]
\begin{minipage}{0.43\textwidth}
\includegraphics[width=\textwidth]{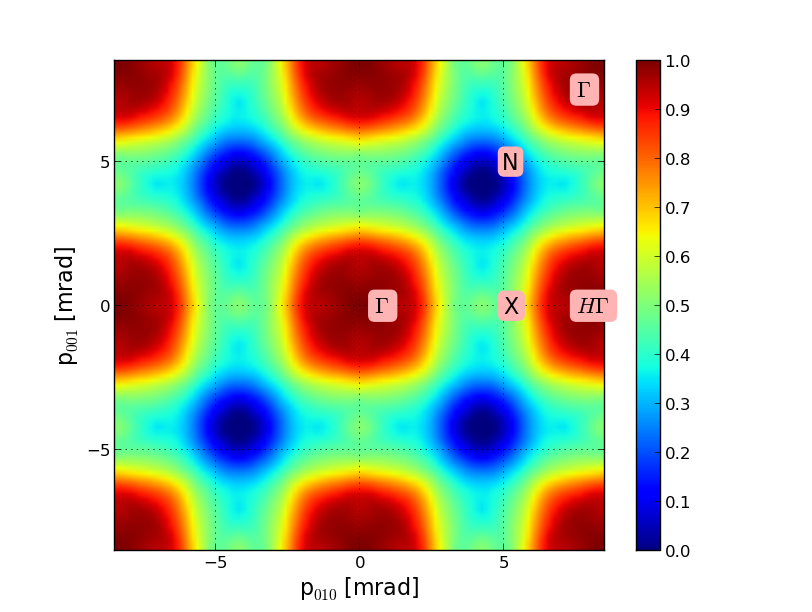}
\caption{\label{fig:rech}LCW distribution for paramagnetic Cr [100] obtained from first principle calculations.}
\end{minipage}\hspace{0.05\textwidth}%
\begin{minipage}{0.43\textwidth}
\includegraphics[width=\textwidth]{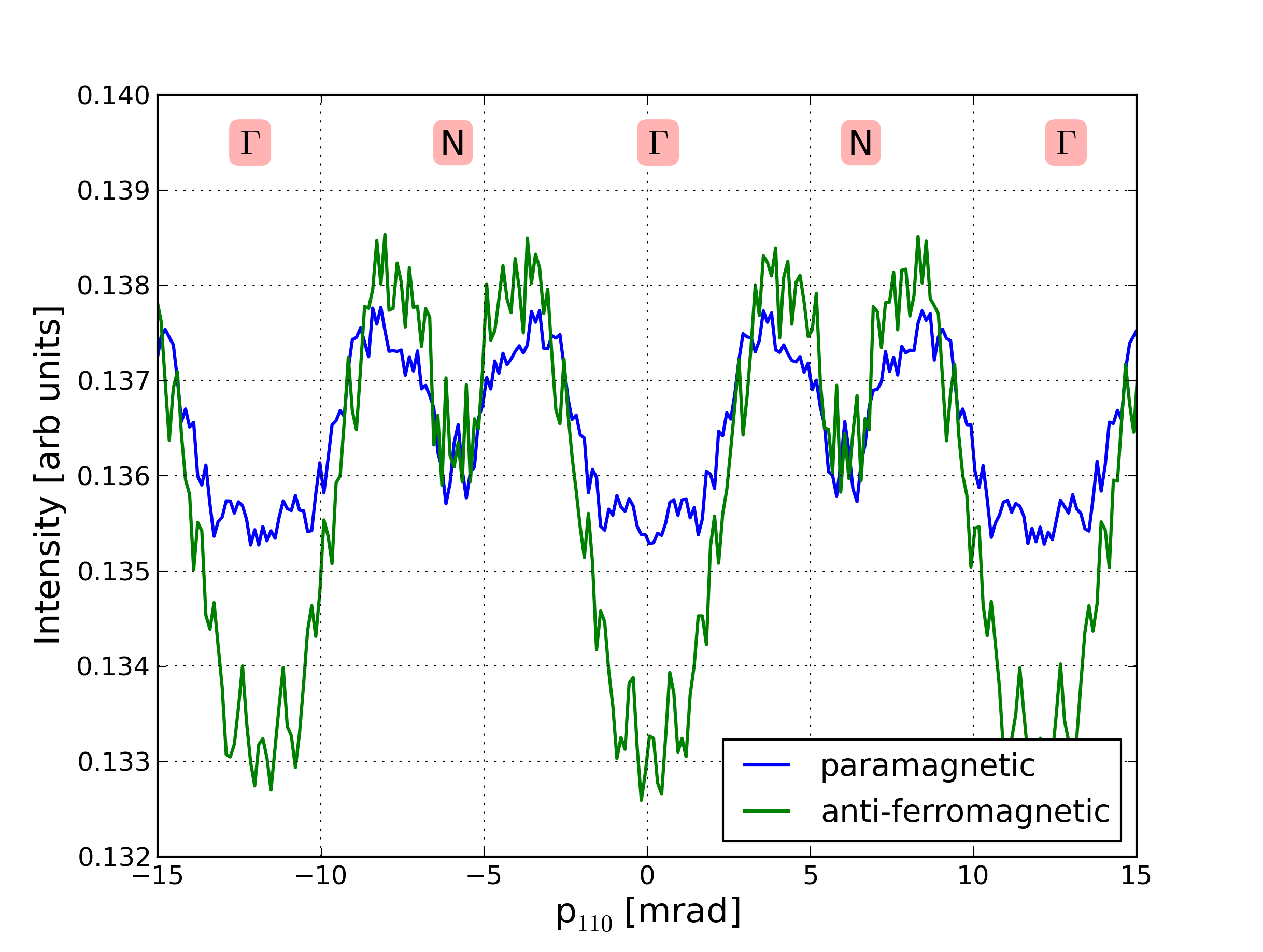}
\caption{\label{fig:cut1}Cut through the LCW-folded projections  along $\Gamma$-N-$\Gamma$ (p$_{001}=$p$_{010}$) for the paramagnetic and the anti-ferromagnetic phase. The initial LCW-projections were normalized to the same amount of counts.}
\end{minipage} 
\end{figure}
Two of the four  measurements were performed at room temperature in order to check if the ACAR spectra are consistent before and after the change of the sample holder. Since both spectra are compatible within the statistical limits only  the spectrum with the higher statistics is presented here (see figure \ref{fig:room}). The comparison of the low temperature measurement  and the room temperature measurement shows the effect of decreasing resolution with increasing temperature. However, besides the lower resolution in the room temperature measurement both spectra are comparable, therefore only one is shown. \\
The features at the N-points appear to be occupied in the anti-ferromagnetic phase (High symmetry points of the Brillouin zones of both phases, \textit{simple cubic} for the anti-ferromagnetic phase and \textit{bcc} for the paramagnetic phase are indicated in figures 1--5). This picture changes when the sample temperature is increased above the Néel temperature (see figure \ref{fig:318}). A small difference in temperature drives the sample into the paramagnetic state and yields a drastic change in the LCW-folded data. The N-hole pockets become more pronounced, which can be seen comparing cuts along $\Gamma$-N-$\Gamma$ for both phases (see figure \ref{fig:cut1}). Furthermore, the electron pocket structure at the X-points connecting the N-points is smaller than in the anti-ferromagnetic phase. Generally, the agreement of our measurement with previous 2D-ACAR measurements (see ref.\cite{Dug}) on paramagnetic Cr is considerable. However, the discrepancies between our data and the SPRKKR calculations are quite  substantial. This has also been reported from different laboratories \cite{Dug,Bia}. The relative intensities in the calculation for the occupied and the unoccupied states can not be reproduced in the LCW-folded data (see figure \ref{fig:rech}). 
\section{Discussion}
\begin{figure}[h]
\includegraphics[width=16pc]{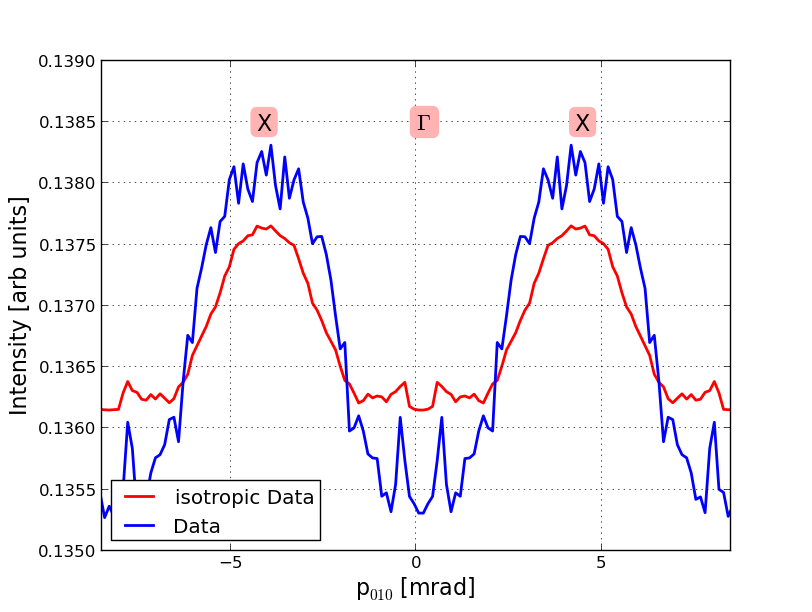}\hspace{2pc}%
\begin{minipage}[b]{14pc}\caption{\label{fig:cut}Cut through the LCW-folded data along $X\Gamma X$ (p$_{001}=0$) in comparison with the same cut for a LCW-folding of the isotropic data. The position of high symmetry points is indicated.}
\end{minipage}
\end{figure}

The experimental resolution of our spectrometer lies between $1.54\,$mrad and $1.64\,$mrad \cite{Hub}, which is only slightly less than the extent of the N-point features in our data. However, the changes in the LCW-folded data when the sample is driven over the magnetic phase transition cannot be attributed to resolution effects, since a temperature increase of $20\,$K results in a degradation of the overall resolution of less than $0.01\,$mrad.\\
 During data analysis the LCW-folding was applied to the isotropic data, which was obtained by azimuthal averaging of the original data. The isotropic data does not contain the full information on the Fermi surface topology. With the isotropic data alone features at the X-point, similar to those described above, can be produced by the LCW-folding, as can be seen in figure \ref{fig:cut}.\\
A possible explanation would be that the positron annihilation with a high lying flat band close to the Fermi energy is strongly enhanced. Also it has to be considered the application of the LCW-theorem is only valid  if the positron wave function can be characterized as a plane wave, which is not true if strong correlation effects are present, like it is the case for Cr \cite{Min}.\\
To summarize, we were able to reproduce the results from previous measurements\cite{Dug} we also could show, that the 2D-ACAR spectrum is drastically different in the anti-ferromagnetic and in the paramagnetic phase. However, the interpretation of our data with respect to the calculation remains an unsolved challenge.
\section{Outlook}
For the interpretation of the  2D-ACAR data of Cr a solid understanding of the enhancement is needed. We plan on recording multiple 2D-ACAR projections of Cr in the paramagnetic phase in order to perform a full 3D reconstruction of the momentum density, respective measurements are under way. This way, we hope to get a better grasp of the enhancement in Cr and with that gain insight to the electronic structure. 
\section*{Acknowledgements}
This project is funded by the Deutsche Forschungsgesellschaft (DFG) within the Transregional Collaborative Research Center TRR 80 ``From electronic correlations to functionality''.


\section*{References}
\begin{thebibliography}{9}
\bibitem{CrRev} Steinitz M O 1986 \textit{J. Magn. Magn. Mater. } \textbf{60} 137-144
\bibitem{Shi} Shiotani N et al. 1977 \textit{J. Phys. Soc. Jpn. } \textbf{43} 4
\bibitem{Sin} Singh A K and Singru R M 1983 \textit{J. Phys. F: Met. Phys.} \textbf{13}  2189-2196
\bibitem{Fret} Hughes R J et al. 2004 \textit{Phys. Rev. B} \textbf{69}  174406
\bibitem{Sint} Singh A K et al. 1988 \textit{Europhys. Lett.} \textbf{6} 67 
\bibitem{Sun} Sundamrajant V, Kanheret D G  and Singru  R M  1992 \textit{J. Phys.: Condens. Matter} \textbf{4}  89754988
\bibitem{Rub} Rubaszek A  et al. 2001  \textit{Phys.Rev. B} \textbf{65} 125104
\bibitem{Dug} Dugdale S B et al. 1998 \textit{J. Phys.: Condens. Matter} \textbf{10} 10367
\bibitem{Bia} Biasini M 2000 \textit{Physica B} \textbf{275} 285-294
\bibitem{Lav} Laverock J et al. 2010 \textit{Phys.Rev. B} \textbf{82} 125127
\bibitem{Hub} Ceeh H et al. 2013 \textit{Rev. Sci. Instrum.} \textbf{84} 043905
\bibitem{LCW} Lock D G, Crisp V H C, West R N 1973 \textit{J. Phys. F: Met. Phys.} \textbf{3} 561
\bibitem{Theo} Ebert H, Kodderitzsch  D, Minár J 2011\textit{ Rep. Prog. Phys.} \textbf{74} 096501
\bibitem{Min} Rabou LP L M   and Mijnarends P E  1984 \textit{ Solid State Commun.} \textbf{52}  933-936

\end{thebibliography}
\end{document}